\documentclass[aps,10pt,prb,twocolumn,showpacs,amsmath,amssymb,superscriptaddress]{revtex4-1}

\usepackage{graphicx}
\usepackage{hyperref}
\hypersetup{
 pdfnewwindow=true, colorlinks=true,
 linkcolor=blue, anchorcolor=blue,
 citecolor=blue, filecolor=blue,
 menucolor=blue, urlcolor=blue}
\usepackage{breakurl}
\usepackage{dcolumn}
\newcolumntype{.}{D{.}{}{1}}

\begin{document}

\title{Canonical magnetic insulators with isotropic magnetoelectric
  coupling}

\author{Sinisa Coh}
\email{sinisa@civet.berkeley.edu} 
\affiliation{Department of Physics, University of California at
  Berkeley and Materials Sciences Division, Lawrence Berkeley National
  Laboratory, Berkeley, California 94720, USA}

\author{David Vanderbilt}
\affiliation{ Department of Physics \& Astronomy, Rutgers University,
  Piscataway, New Jersey 08854, USA}

\date{\today}


\begin{abstract} 
  We have performed an exhaustive representation-theory-based search
  for the simplest structures allowing isotropic magnetoelectric
  coupling.  We find 30 such structures, all sharing a common pattern
  of atomic displacements in the direction of atomic magnetic moments.
  We focus on one of these 30 canonical structures and find that it is
  generically realized in a class of fractionally substituted
  pyrochlore compounds with an all-in-all-out magnetic order.
  Furthermore, we find that these substituted pyrochlore compounds
  have a substantial Chern-Simons orbital magnetoelectric component
  ($\theta=0.1$~--~$0.2$).  While this component is also formally
  present in strong Z$_2$ topological insulators ($\theta=\pi$), its
  effects are observable there only if time-reversal symmetry is
  broken at the surface.
\end{abstract}

\maketitle

One of the characteristics of the interplay between electric and magnetic
degrees of freedom is a linear magnetoelectric tensor $\alpha_{ij}$.
It expresses the {\it electric} polarization $P_i$ induced in an insulator
by an applied {\it magnetic} field $B_j$. Such a response requires broken
time-reversal and inversion symmetries, and is known to occur in some
compounds such as Cr$_2$O$_3$. In general, the tensor $\alpha_{ij} =
\partial P_i / \partial B_j$ has nine independent coefficients.  From the
symmetry point of view the simplest possible tensor $\alpha_{ij}$ is
diagonal, with all elements on the diagonal being equal,
\begin{align}
  \alpha_{ij} = \alpha^{\rm iso} \delta_{ij} =
  \left(
    \begin{array}{...}
      \alpha.^{\rm iso} & 0. & 0. \\
      0. & \alpha.^{\rm iso} & 0. \\
      0. & 0. & \alpha.^{\rm iso}
    \end{array}
  \right).
  \label{eq:iso}
\end{align}
Materials with such an isotropic magnetoelectric (ME) response have
been discussed in the literature,\cite{meipic5, hehl09,khomskii} but
to our knowledge no such materials have yet been reported
experimentally.

Recently, the interest in isotropic ME response has grown sharply due
to the discovery\cite{qi08,essin09} of a mechanism giving rise to a
purely isotropic ME coefficient, and relationship of this finding to
the physics of strong Z$_2$ topological
insulators.\cite{FuPrl2007,MoorePrb2007,qi08,essin09} This component
$\alpha^{\rm CS}$ of the ME response is referred to as the
Chern-Simons orbital ME polarizability (CSOMP) and is conventionally
measured in terms of the dimensionless parameter $\theta$ via
\begin{align}
  \alpha^{\rm CS}_{ij}=\theta \frac{e^2}{2\pi h}  \delta_{ij}.
  \label{eq:CS}
\end{align}
Here $e$ is the electron charge and $h$ is Planck's constant. In what
follows we denote the entire isotropic ME response as $\alpha^{\rm
  iso}$, and its Chern-Simons component as $\alpha^{\rm CS}$.  In
strong Z$_2$ topological insulators, formally $\theta$\,=\,$\pi$ and
$\alpha$\,=\,$\alpha^{\rm iso}$\,=\,$\alpha^{\rm CS}$\,=\,$e^2/2h$,
but this ME coupling is observable only if the surfaces and interfaces
of the sample are consistently gapped by some time-reversal-breaking
perturbation.\cite{Qi2009, essin09, Coh2011} In Cr$_2$O$_3$ and other
conventional magnetic insulators, on the other hand, $\alpha$ is
easily observable but $\alpha^{\rm CS}$ is small.  We seek here a
materials where $\alpha^{\rm CS}$ is both large and observable, as
might be the case in a magnetic insulator that is close to being a
strong Z$_2$ topological insulator.\cite{Coh2011}

A critical consideration which determines which of the nine components
of $\alpha_{ij}$ can be non-zero is that of symmetry.\footnote{In
  fact, symmetry considerations were important in the study of
  magnetoelectrics from the early beginnings.\cite{Curie1894}} In many
known magnetic insulators, symmetry allows only off-diagonal
components of $\alpha_{ij}$ to be non-zero, as for example in
Li(Fe,Co,Ni)PO$_4$ and many other compounds which have only
$\alpha_{xy}$ and $\alpha_{yx}$ different from zero.  Similarly, the
series of compounds (Tb,Dy,Ho)PO$_4$ have $\alpha_{xx}=-\alpha_{yy}$
as the only two non-zero components. On the other hand, some compounds
such as Cr$_2$O$_3$ have a diagonal $\alpha$ of the form
$\alpha_{xx}=\alpha_{yy}\neq\alpha_{zz}$, which is only isotropic if
artificially tuned to be so.  However, in Cr$_2$O$_3$ one expects the
response along the rhombohedral axis $\alpha_{zz}$ to arise from a
different microscopic mechanism than the $\alpha_{xx}$ and
$\alpha_{yy}$ components, since the spin moments on the Cr atoms are
aligned along the $\pm$z direction and can easily tilt towards the x-y
plane.\cite{Jorge2008,Malashevich2012}

The three main contributions of this work are as follows. First, we
find an exhaustive list of the 30 simplest crystal structures and
corresponding arrangements of magnetic moments which, by symmetry,
allow a purely isotropic linear ME coupling $\alpha^{\rm
  iso}=\alpha_{xx}=\alpha_{yy} =\alpha_{zz}$.  Second, using
density-functional theory calculations we find that one of these 30
cases is generically realized in any member of a class of substituted
pyrochlore compounds with all-in-all-out magnetic order.  Third, we
find a relatively large CSOMP component ($\alpha^{\rm
  CS}$=0.85--1.62\,ps/m) from our calculations on these substituted
pyrochlore compounds.

We start with an analysis of the required symmetry breaking which
would allow for an isotropic linear ME coupling.  Let us first
consider the effects of symmetry operations on the isotropic ME
coupling coefficient $\alpha^{\rm iso}$ defined in Eq.~(\ref{eq:iso}).
The real number $\alpha^{\rm iso}$ must change sign both under the
time-reversal transformation (since it transforms $B_j$ into $-B_j$)
and under the inversion transformation (since it transforms $P_i$ into
$-P_i$).  Furthermore, $\alpha^{\rm iso}$ is unchanged under rotation
(since it measures an isotropic response) or translation (since it is
a bulk response).  Therefore, there are two classes of transformations
which change the sign of $\alpha^{\rm iso}$.  The first class (class
T) of transformations consists of the time-reversal operator either by
itself or followed by a proper rotation and/or a translation.  The
second class (class P) of transformations consists of the inversion
symmetry either by itself or followed by a proper rotation and/or a
translation.  Therefore, any system with at least one symmetry
operation in either class T or P is excluded as a candidate for an
isotropic ME material.

Consider the example of a simple cubic (primitive) lattice with one
atom per unit cell, and a magnetic moment on that atom pointing along
the $z$-axis. Such a system has broken time-reversal symmetry.
However, time-reversal symmetry followed by a 2-fold rotation around
the $x$-axis will still be a symmetry (class T), and it will enforce
$\alpha^{\rm iso}=0$. (In fact, in this simple example, even inversion
symmetry would enforce $\alpha^{\rm iso}=0$.) Therefore, as mentioned
earlier, it is clear that not every magnetic order has the correct
symmetry to produce a non-zero isotropic ME coupling.

We now search for the highest-symmetry atomic structures with the
property that the isotropic linear ME coupling ($\alpha^{\rm iso}$) is
allowed by symmetry, i.e., no symmetry elements of the system belong
to either class T or P.  In other words, such a structure has only
those symmetry-breaking perturbations which are essential to allow
$\alpha^{\rm iso} \neq 0$.  We start our search by selecting a set of
the simplest periodic arrangements of atoms (ignoring their magnetic
moments for now).  We consider all 36 space groups in the cubic system
($T$, $T_h$, $T_d$, $O$, and $O_h$ point groups) and their 308 Wyckoff
orbits (symmetry-related subsets of atoms).  For simplicity we only
consider Wyckoff orbits with at most one free parameter (227 out of
308). Furthermore, we only consider characteristic\cite{Engel1984}
orbits (whose symmetry is not larger than that of an underlying space
group), leaving us with 62 orbits out of 227.

We now take each of these 62 highest-symmetry periodic arrangements of
atoms, and consider all possible arrangements of atomic magnetic
moments that do not enlarge the chemical unit cell. Such an
arrangement of magnetic moments is described by $N$ magnetic-moment
vectors, or equivalently, $3N$ Cartesian variables, where $N$ is the
number of sites in the Wyckoff orbit.  We construct a $3N \times 3N$
matrix for each symmetry operator in the space group, taking into
account the axial nature of the magnetic moment (no sign change under
inversion). Using standard space-group character tables we can
decompose these $3N \times 3N$ matrices into irreducible
representations.\cite{Bertaut1968} Next, we consider all
one-dimensional real irreducible representations which satisfy our
symmetry constraint, namely, the characters are $+1$ or $-1$ for
symmetry operators with positive or negative determinant (proper or
improper rotations) respectively.  In other words, such an arrangement
of magnetic moments is symmetric under proper rotations, and is
symmetric under improper rotations only when coupled with a
time-reversal operation.

As a result of this search, we find that 30 of the 62 structures yield
a unique periodic arrangement of atoms with magnetic moments
satisfying our constraints, while the remainder yield none.  These 30
are the simplest (canonical) arrangements of atoms and corresponding
magnetic moments allowing for a purely isotropic linear ME coupling.
All 30 of these arrangements are listed in the Supplement of this
manuscript.\cite{supp}

Analyzing these 30 cases, we find two common features. First, we find
a {\it local} motif in which atoms are displaced away from
higher-symmetry locations (as described by a single free Wyckoff
parameter) and magnetic moments point in the same direction as the
displacements.  Second, we find that this local motif is arranged in
an appropriate three-dimensional network of polyhedra ({\it global}
consistency). Finally, we find that the atomic displacements cause
breathing of these polyhedral networks so as to increase the size of
half of the polyhedra and decreases the size of the other half.

\begin{figure}[!t]
\centering\includegraphics{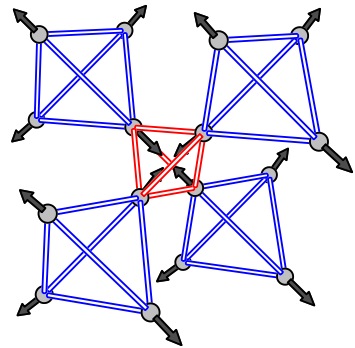}
\caption{Sketch of one out of 30 canonical structures in which
  isotropic ME coupling $\alpha^{\rm iso}$ is allowed by
  symmetry.  Atoms (gray spheres) are displaced in the direction of
  their magnetic moments (arrows). These displacements change the
  network of corner sharing tetrahedra by shrinking half of 
  the tetrahedra (red) and enlarging the other half (blue).}
\label{fig:3d}
\end{figure}

The magnetic part of this motif, with moments pointing all-in and
all-out in neighboring polyhedra, is not very difficult to realize in
nature.  For example, such a magnetic order has been suggested in a
variety of pyrochlores with magnetic Os, Ir, or rare-earth atoms,
and was first predicted theoretically in Ref.~\onlinecite{Xiangang2011}.
A well-known example is Cd$_2$Os$_2$O$_7$, which is experimentally found
to be consistent with a long-range all-in-all-out magnetic order on
the Os site as shown in Ref.~\onlinecite{Sleight1974,Yamaura2012} and
in theoretical calculations in Ref.~\onlinecite{Shinaoka2012}.
Similarly, all-in-all-out magnetic order occurs on the Ir site in
Eu$_2$Ir$_2$O$_7$ as shown in Ref.~\onlinecite{Sagayama2013} and on
both Nd and Ir sites in Nd$_2$Ir$_2$O$_7$ as shown in
Ref.~\onlinecite{Tomiyasu2012}.  Thus, we may achieve the desired
isotropic magnetoelectric coupling if we can augment the observed
noncollinear spin ordering with an appropriately similar pattern of
atomic displacements.

With this motivation, in the remainder of this paper we focus on just
one of our 30 structures, namely, the one having space group
F$\bar{4}$3m, Wyckoff orbit $16e$, and magnetic moments corresponding
to the irreducible representation $\Gamma_2$ (structure 13 in the
Supplement\cite{supp}).  The coordinates\footnote{In reduced
  coordinates of the conventional face-centered cubic cell.}  of the
atoms in this Wyckoff orbit are $(u,u,u)$, $(u,-u,-u)$, $(-u,u,-u)$,
and $(-u,-u,u)$, with an arbitrary value of the real number parameter
$u$. The directions of the magnetic moments in the $\Gamma_2$
representation are $(1,1,1)$, $(1,-1,-1)$, $(-1,1,-1)$, and
$(-1,-1,1)$ respectively.  In the special case that $u=1/8$ (or
equivalently $3/8$, $5/8$, or $7/8$) the symmetry increases from
F$\bar{4}$3m to Fd$\bar{3}$m, the Wyckoff orbit notation changes to
$16c$ or $16d$, and the ME coupling is forced to zero by symmetry.
Such an arrangement corresponds to a network of corner-sharing
tetrahedra as formed for example by the A or B sites of the
pyrochlores A$_2$B$_2$O$_7$ or the B sites of spinels AB$_2$O$_4$.
Displacing atoms away from $u=1/8$ by taking $u=1/8 + \epsilon$ for
some small $\epsilon$ leads to a breathing distortion of the
tetrahedral network, with half of the tetrahedra increasing in size
and the other half shrinking (see red versus blue tetrahedra in
Fig.~\ref{fig:3d}). The direction of each atomic displacement is the
same as the direction of the corresponding local magnetic moment in
the $\Gamma_2$ representation (so called all-in-all-out magnetic
arrangement shown by arrows in Fig.~\ref{fig:3d}).

We now focus on a realization of this particular canonical structure
in a substituted A$_2$B$_2$O$_7$ pyrochlore. The compositional formula
of the pyrochlores is often written as A$_2$B$_2$O$_6$O$'$ since
oxygen atoms occupy two distinct crystallographic sites labeled O and
O$'$, with the O$'$ sites centered inside the tetrahedra formed by the
A lattice.  Based on a symmetry analysis of the pyrochlore lattice, we
find that if the O$'$ sites are divided into two regular sublattices,
one of which is substituted by a different atom (or a vacancy), the
symmetry of the pyrochlore is reduced to F$\bar{4}$3m, with both A and
B atoms moved onto $16e$ Wyckoff orbits.  (This does not double the
size of the primitive cell.)  Therefore, we generically expect that
any A$_2$B$_2$O$_7$ pyrochlore with all-in-all-out magnetic order on
the A sites, the B sites, or both, will become an isotropic ME upon
50\% O$'$ substitution.

\begin{figure}[!t]
\centering\includegraphics{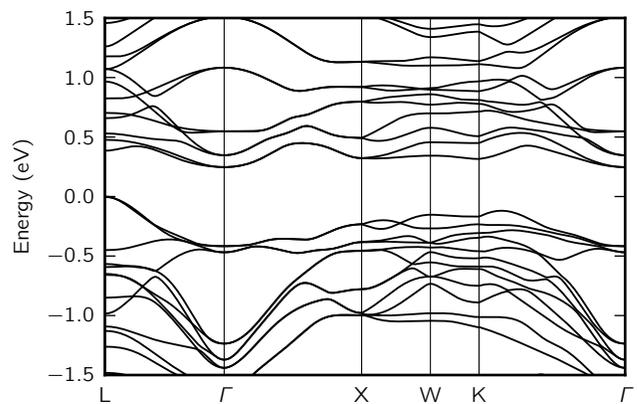}
\caption{First-principles fully relativistic computed
  band structure of Zn$_2$Os$_2$O$_7$ with partial Te substitution
  (Zn$_2$Os$_2$O$_{6.5}$Te$_{0.5}$).}
\label{fig:bs}
\end{figure}

Indeed, we confirm these findings using first-principles calculations
on two families of pyrochlore compounds.  The first family we analyzed
have non-magnetic A$^{2+}$ ions and magnetic B$^{5+}$ ions.  We find
that these compound have an isotropic ME coupling for any combination
of A = (Cd, Zn, Hg), B = (Os, Ru) and X = (S, Se, or Te) substituting
half of the O$'$ sites.  (The general formula of such a compound is
A$_2$B$_2$O$_{6.5}$X$_{0.5}$.)  Using methods from
Ref.~\onlinecite{Coh2011}, we have computed $\theta$ for
Zn$_2$Os$_2$O$_{6.5}$Te$_{0.5}$, using PBE
exchange-correlation\cite{perdew-prl96} and an onsite Hubbard $U$ of
3.0~eV\cite{Liechtenstein1995} (we find negligibly small dependence of
$\theta$ on $U$).  We obtain $\theta=$0.11, corresponding to
$\alpha^{\rm CS}$=0.85~ps/m.\footnote{As an order-of-magnitude
  comparison, the entire (spin-ion dominated) magnetoelectric response
  in the prototypical magnetoelectric Cr$_2$O$_3$ is about
  1.04~ps/m.\cite{Malashevich2012}} The band structure of this
compound is shown in Fig.~\ref{fig:bs}; the minimum direct and
indirect band gaps are 0.39~eV and 0.25~eV respectively, and the Os
magnetic moment is 0.94~$\mu_{\rm B}$.

We also find that replacing B=Os with B=Ru roughly doubles $\theta$
($\theta$=0.21 in Cd$_2$Ru$_2$O$_{6.5}$Te$_{0.5}$, corresponding to
$\alpha^{\rm CS}$=1.62~ps/m).  Unfortunately, B=Ru compounds tend to
become semi-metallic upon Se or Te substitution, at least within the
density-functional approximation (here PBE), which often
underestimates gaps.  However these compounds may be easier to handle
experimentally since they do not contain toxic Os.  For this first
family we find, in general, that swapping the A site from Zn to Cd to
Hg somewhat reduces $\theta$, while swapping X from S to Se to Te
increases $\theta$.

The second family of pyrochlore compounds that we have analyzed have
magnetic ions on both A and B sites, typically rare-earth A$^{3+}$
ions and other B$^{4+}$ ions.  Since both ions are magnetic, one might
expect larger values of $\theta$ than in the first family.
Furthermore, magnetism on the A sites is preferred over magnetism on
the B sites, since the A site is about two times closer (2.2~\AA\
versus 4.2~\AA) to the substituted atom X than the B atom. We have not
attempted to calculate the values of $\theta$ in these compounds,
since it is well known that local DFT approximations do not treat $f$
valence electrons reliably in the rare-earth atoms.

In order to relate our predictions to previous experimental work, we
give a brief overview here of known pyrochlore compounds with
breathing distortions that can potentially be combined with
all-in-all-out magnetic order.  We are unaware of any other prediction
of a pyrochlore compound in which breathing and magnetism occur at the
same time (and the system remains insulating).  This is consistent
with the spirit of Ref.~\onlinecite{Hill2000}, where it was argued
that atomic distortions and magnetic moments tend not to happen on the
same atomic site.  (At least, this was argued for oxides; it is
unclear whether it should also hold for fluorides.\cite{Scott2011})
Our proposed mechanism for displacing the magnetic ions in the
pyrochlores does not, however, rely on any intrinsic displacive
tendency; it is more reminiscent of the case of BiFeO$_3$, where an
independent mechanism -- Bi off-centering there, O$'$ replacement here
-- provides the driving force for magnetic-ion off-centering.

The substitution of the O$'$ site with sulfur has been analyzed
previously in Ref.~\onlinecite{Bernard1973} (and reviewed in
Ref.~\onlinecite{SubramanianReview1983}) in Cd$_2$Nb$_2$O$_{7-x}$S$_x$
for the entire range $0<x<1$ of substitution.  While the arguments
given above were for the case of a full 50\% substitution (complete
replacement of a sublattice), note that the same effects will occur,
and a nonzero $\alpha^{\rm iso}$ will be generated, if only a partial
replacement is carried out by alloying on the O$'$ site, as long as
the concentration is different for the two O$'$ sublattices.  Only
polycrystalline Cd$_2$Nb$_2$O$_{7-x}$S$_x$ samples were made, and it
is not known whether substituted atoms are ordered or not. However,
somewhat suggestive of an ordered state is a
finding\cite{Bernard1973,SubramanianReview1983} that its structural
phase diagram differs for $x<0.5$ compared to $x>0.5$, with the
$x>0.5$ case being more complicated. More detailed structural studies
have been made on the somewhat related compounds Pb$_2$Ir$_2$O$_{6.5}$
in Ref.~\onlinecite{Kennedy1996} and Pb$_2$Ru$_2$O$_{6.5}$ in
Ref.~\onlinecite{Beyerlein1984}. In these compounds, instead of
substitution, half of the O$'$ sites are replaced with vacancies.
Detailed structural studies in these compounds show that vacancies are
indeed in the long-range-ordered arrangement.  \footnote{However,
  these compounds are metallic due to vacancies, and therefore are not
  interesting candidates for magnetoelectrics.  Nevertheless,
  counter-doping on either A or B site could possibly make these
  systems insulating.}
Even if the compositional ordering is only short-ranged, we point out
that the commonly-used ME annealing technique can rearrange ME domains
in the sample so that each locally ordered region has the same sign of
$\alpha^{\rm iso}$.  Therefore, it may be enough to require that
A$_2$B$_2$O$_{6.5}$X$_{0.5}$ has only locally-ordered substituted
atoms X.

We leave for future work the analysis of other realizations of this
particular canonical structure (Wyckoff orbit $16e$ in F$\bar{4}$3m
can also appear in spinels), as well as the study of the remaining 29
(out of 30) canonical magnetically-decorated structures. Nevertheless,
we point out some interesting candidates among these.  For example,
chromium boracites Cr$_3$B$_7$O$_{13}$Br and Cr$_3$B$_7$O$_{13}$I are
believed to have an anti-ferromagnetic ground state\cite{schnelle1999}
and may be magnetoelectric.\cite{meipic5} Symmetry lowering of the Cr
atoms to Wyckoff orbit 24$g$ in group F23 would allow $\alpha^{\rm
  iso} \neq 0$ (the corresponding characteristic orbit is 24$e$ in
Fm$\bar{3}$m). The ullmannite structure\cite{Foecker2001} found in
NiSbS (and many other compounds) is also interesting, as it consists
of three different atoms on the 4$a$ orbit in the space group P2$_1$3.
Similarly, silicides such as FeSi and MnSi are composed of the same
Wyckoff orbit (4$a$ orbit in P2$_1$3) and are known to be magnetic.

In summary, our work shows that pyrochlores partially substituted by
S, Se, or Te, and having magnetic all-in-all-out magnetic order,
generically have a nonzero and purely isotropic linear ME coupling.
If such a compound is made in the laboratory, and its ME coupling is
measured, this would be the first realization of a purely isotropic ME
material, and also a first known material with substantial\footnote{In
  a prototypical magnetoelectric\cite{Malashevich2012} such as
  Cr$_2$O$_3$, the orbital-electronic component is responsible for
  only 1\% of the entire ME response. Furthermore, this 1\% is mostly
  dominated by Kubo-like terms, while the purely isotropic
  (Chern-Simons) $\theta$-component is only about 0.1\% of the full
  response.}  ME response resulting from the orbital-electronic
mechanism.\cite{Malashevich2012} For example, a value of $\theta=0.1$,
which was shown above to be quite plausible, would make the orbital
magnetoelectic coupling comparable to the entire (spin-ion dominated)
response in Cr$_2$O$_3$.  Furthermore, such a compound would have
substantial Chern-Simons ME polarizability, which is formally present
in strong Z$_2$ topological insulators as well, but is hidden from
observation.  Finally, we believe that similar exhaustive
symmetry-based analyses could be used in the search for topological
superconductors, Weyl semi-metals, or other magnetoelectric and/or
multiferroic classes of materials.

\begin{acknowledgments}
  We acknowledge discussion with S-W. Cheong, D. G. Schlom, and Weida
  Wu.  S.C.\ acknowledges support by the Director, Office of Energy
  Research, Office of Basic Energy Sciences, Materials Sciences and
  Engineering Division, of the U.S. Department of Energy under
  contract DE-AC02-05CH11231 which provided for the density functional
  theory calculations.  D.V.\ acknowledges support from NSF Grant
  DMR-10-05838.
\end{acknowledgments}

\bibliography{pap}

\end{document}